\documentclass[12pt]{article}

\newcommand{\beq}{\begin{equation}}
\newcommand{\eeq}{\end{equation}}
\newcommand{\beqa}{\begin{eqnarray}}
\newcommand{\eeqa}{\end{eqnarray}}
\newcommand{\beqar}{\begin{eqnarray*}}
\newcommand{\eeqar}{\end{eqnarray*}}

\newcommand{\eps}{\epsilon}
\newcommand{\ga}{\gamma}
\newcommand{\Ga}{\Gamma}

\newcommand{\inn}{\!\cdot\!}

\newcommand{\z}{\zeta}

\newcommand{\eg}{{\it e.g.,}\ }
\newcommand{\ie}{{\it i.e.,}\ }
\newcommand{\labell}[1]{\label{#1}} 
\newcommand{\reef}[1]{(\ref{#1})}
\newcommand\prt{\partial}
\newcommand\veps{\varepsilon}

\newcommand\cR{{\cal R}}

\newcommand\cF{{\cal F}}
\newcommand\cM{{\cal M}}
\newcommand\cN{{\cal N}}
\newcommand\cB{{\cal B}}

\newcommand\hR{\hat{R}}

\newcommand\Tr{{\rm Tr}}
\newcommand\tr{{\rm tr}}

\begin{document}

\begin{titlepage}

\begin{center}



\vskip 2 cm
{\LARGE \bf S-duality of D-brane action\\  at order $O(\alpha'^2)$
 }\\
\vskip 1.25 cm
 Mohammad R. Garousi\footnote{garousi@ferdowsi.um.ac.ir}  \\
 \vskip 1cm
\vskip 1 cm
{{\it Department of Physics, Ferdowsi University of Mashhad\\}{\it P.O. Box 1436, Mashhad, Iran}\\}
\vskip .1 cm
\vskip .1 cm

\end{center}

\vskip 0.5 cm

\begin{abstract}
\baselineskip=18pt
Using the compatibility of   the DBI and the Chern-Simons actions with  the T-duality transformations, the curvature corrections to these actions  have been recently extended  to include the quadratic B-field couplings at order $O(\alpha'^2)$.  In this paper, we use  the compatibility of the couplings on D$_3$-brane with the S-duality to  find the nonlinear RR  couplings at order $O(\alpha'^2)$.   We confirm  the  quadratic RR couplings in the DBI part with the disk-level scattering amplitude. Using the regularized non-holomorphic Eisenstein series $E_1$,  
 we then write the results  in  $SL(2,Z)$ invariant form.

\end{abstract}
\vskip 3 cm
\begin{center}
\end{center}
\end{titlepage}
\section{Introduction}
The  low energy effective field theory of  D-branes  consists of the  Dirac-Born-Infeld (DBI) \cite{Bachas:1995kx} and the  Chern-Simons (CS) actions \cite{Douglas:1995bn}. The curvature corrections to the CS part  can be found by requiring that the chiral anomaly on the world volume of intersecting D-branes (I-brane) cancels with the anomalous variation of the CS action. This  action for a single D$_p$-brane at order $O(\alpha'^2)$ is given by \cite{Green:1996dd,Cheung:1997az,Minasian:1997mm},
\beqa
S_{_{CS}}&\supset&-\frac{\pi^2\alpha'^2T_{p}}{24}\int_{M^{p+1}}C^{(p-3)}\wedge\bigg[\tr (R_T\wedge R_T)-\tr (R_N\wedge R_N)\bigg]\labell{CS}
\eeqa
where $M^{p+1}$ represents the world volume of the D$_p$-brane. 
For totally-geodesic embeddings of world-volume in the ambient spacetime,   ${\rm R}_{T,N}$ are the pulled back curvature 2-forms of the tangent and normal bundles respectively (see the appendix in ref.
\cite{Bachas:1999um} for more details).

 The curvature corrections to the DBI  action has been found in \cite{Bachas:1999um} by requiring consistency of the effective action with the $O(\alpha'^2)$ terms of the corresponding disk-level scattering amplitude \cite{Garousi:1996ad,Hashimoto:1996kf}. For totally-geodesic embeddings of world-volume in the ambient spacetime, the corrections in the string frame  for zero B-field and for constant dilaton  are \cite{Bachas:1999um}
\beqa
S_{_{DBI}}&\!\!\!\!\supset\!\!\!\!&\frac{\pi^2\alpha'^2T_{p}}{48}\int d^{p+1}x\,e^{-\phi}\sqrt{-G}\bigg[R_{abcd}R^{abcd}-2\hR_{ab}\hR^{ab}-R_{abij}R^{abij}+2\hR_{ij}\hR^{ij}\bigg]\labell{DBI}
\eeqa
where $\hR_{ab}=G^{cd}R_{cadb}$ and $\hR_{ij}=G^{cd}R_{cidj}$. Here also a tensor with the world-volume or transverse space indices is the pulled back of the corresponding  bulk tensor onto world-volume or transverse space\footnote{Our index conversion is that the Greek letters  $(\mu,\nu,\cdots)$ are  the indices of the space-time coordinates, the Latin letters $(a,d,c,\cdots)$ are the world-volume indices and the letters $(i,j,k,\cdots)$ are the normal bundle indices.}. For the case of D$_3$-brane with trivial normal bundle the curvature couplings \reef{CS} and \reef{DBI} have been modified in \cite{Bachas:1999um} to include the complete sum of D-instanton corrections by requiring  them to  cancel the $SL(2,Z)$ anomalies  of the massless modes of the D$_3$-brane.

In the presence of non-constant dilaton, the couplings \reef{DBI} are not consistent with T-duality. For zero B-field, the compatibility with linear T-duality requires the following extension of \reef{DBI}:
\beqa
S_{_{DBI}}&\!\!\!\!\supset\!\!\!\!&\frac{\pi^2\alpha'^2T_{p}}{48}\int d^{p+1}x\,e^{-\phi}\sqrt{-G}\bigg[R_{abcd}R^{abcd}-2(\hR_{ab}-\phi_{,ab})(\hR^{ab}-\phi^{,ab})\nonumber\\
&&\qquad\qquad\qquad\qquad\qquad\qquad-R_{abij}R^{abij}+2(\hR_{ij}-\phi_{,ij})(\hR^{ij}-\phi^{,ij})\bigg]\labell{DBI2}
\eeqa
where commas denote  partial differentiation\footnote{Using on-shell relations, the definition of the curvature tensor $\hat{R}_{\mu\nu}$ in \cite{Garousi:2009dj} has been changed as $\hat{R}_{\mu\nu}\equiv \frac{1}{2}(R_{\mu a}{}^a{}_{\nu}-R_{\mu k}{}^k{}_{\nu})$. With this tensor the couplings in \reef{DBI}  are then invariant under linear T-duality \cite{Garousi:2009dj} when B-field is zero. If one uses the standard definition $\hat{R}_{\mu\nu}\equiv R_{\mu a}{}^a{}_{\nu}$, then the couplings \reef{DBI2} are invariant.}. The dilaton couplings survive in the Einstein frame even for D$_3$-brane, hence, they break the S-duality of the couplings   found in \cite{Bachas:1999um}. Using the fact that the dilaton and the RR zero-form transform similarly under the S-duality transformations, one expects there should be similar couplings   for the RR zero-form. We will show that the S-matrix element of two RR vertex operators found in \cite{Garousi:1996ad} produces in fact such couplings. Having similar couplings for the dilaton and the RR zero-form, we use an appropriate $SL(2,R)$ matrix to write the results in $SL(2,R)$ invariant form. However,  there is an overall factor of $e^{-\phi}$ in the Einstein frame which is not invariant under the $SL(2,R)$ transformation. 

In the presence of non-zero B-field, the  couplings \reef{CS} and \reef{DBI2} are not consistent with the T-duality. Using the compatibility of  these couplings  with linear T-duality as a guiding principle, the quadratic B-field couplings at order $O(\alpha'^2)$ have been found  in \cite{Garousi:2009dj,Becker:2010ij,Garousi:2010rn}. The B-field couplings in the DBI part are \cite{Garousi:2009dj}\footnote{In this paper, we are interested only in the quadratic order of  field strengths, \eg we are not considering $H^4$,  $RH^2$, or $R\prt\phi\prt\phi$ terms in DBI action.}
\beqa
S_{_{DBI}}&\!\!\!\!\supset\!\!\!\!&-\frac{\pi^2\alpha'^2T_{p}}{48}\int d^{p+1}xe^{-\phi}\sqrt{-G}\bigg[\frac{1}{6}H_{ijk,a}H^{ijk,a}+\frac{1}{3}H_{abc,i}H^{abc,i}-\frac{1}{2}H_{bci,a}H^{bci,a}
\bigg]\labell{LDBI}
\eeqa
 The above couplings have been confirmed with the disk level S-matrix calculations \cite{Garousi:2009dj}. These couplings for D$_3$-brane again break the S-duality of the couplings found in \cite{Bachas:1999um}. Using the fact that the B-field and the RR two-form appear as a doublet in the S-duality transformations, one expects there should be similar couplings as above for the RR two-form. We will show that the S-matrix element of two RR vertex operators found in \cite{Garousi:1996ad} produces in fact such couplings. Having similar couplings for the B-field and the RR two-form, we then write the results in $SL(2,R)$ invariant form. In this case also,  there is an overall factor of $e^{-\phi}$ in the Einstein frame which is not invariant under the $SL(2,R)$ transformation. 

It has been shown in \cite{Garousi:2010ki} that the CS part should  include couplings  which involve linear  NSNS field.  These couplings have been found  by studying the S-matrix element of one RR and one NSNS vertex operators  at order $O(\alpha'^2)$ \cite{Garousi:1996ad}.  These couplings   in the string frame are \cite{Garousi:2010ki}\footnote{Using on-shell relations, the definition of the curvature tensor $\hat{R}_{ij}$ in \cite{Garousi:2010ki} has been changed as $\hat{R}_{ij}\equiv \frac{1}{2}(R_{ia}{}^a{}_j-R_{ik}{}^k{}_j)$. With this tensor the coupling $F^{(p+2)}_{a_0\cdots a_pj,i}\hat{R}^{ij}$ is then invariant under linear T-duality \cite{Garousi:2010ki}. If one uses the standard definition $\hat{R}_{ij}\equiv R_{ia}{}^a{}_j$, then the second term in the second line in \reef{LCS} can be written at the linear order as $F^{(p+2)}_{a_0\cdots a_p}{}^{ j,i}(h_{ij,aa}+h_{aa,ij}-h_{ia,aj}-h_{ja,ai}-2\phi_{,ij})/2(p+1)$ where $h$ is the metric perturbation. Under T-duality along the world volume direction  $y$, the RR factor $F^{(p+2)}_{a_0\cdots a_p}{}^{ j,i}/(p+1)$ which includes the Killing index $y$, transforms to $F^{(p+1)}_{a_0\cdots a_{p-1}}{}^{ j,i}$. This RR field, however,  does not include the Killing index. Hence, the indices $i,j$ in the T-dual theory do not include the Killing index $y$. Using this observation, one can easily verify  that the metric/dilaton factor $(h_{ij,aa}+h_{aa,ij}-h_{ia,aj}-h_{ja,ai}-2\phi_{,ij})$ is invariant under the linear T-duality. Hence, the second term in the second line in \reef{LCS} is invariant under the T-duality.} 
\beqa
S_{_{CS}}&\!\!\!\supset\!\!\!&-\frac{\pi^2\alpha'^2T_p}{24}\int d^{p+1}x\,\eps^{a_0\cdots a_p}\left(\frac{1}{2!(p-1)!}[{ F}^{(p)}_{ia_2\cdots a_p,a}H_{a_0a_1}{}^{a,i}-{ F}^{(p)}_{aa_2\cdots a_p,i}H_{a_0a_1}{}^{i,a}]\right.\labell{LCS}\\
&&\left.\qquad\qquad\qquad\qquad+\frac{2}{p!}\bigg[\frac{1}{2!}F^{(p+2)}_{ia_1\cdots a_pj,a}R^a{}_{a_0}{}^{ij}-\frac{1}{p+1}F^{(p+2)}_{a_0\cdots a_pj,i}(\hat{R}^{ij}-\phi^{,ij})\bigg]\right.\nonumber\\
&&\left.\qquad\qquad\qquad\qquad-\frac{1}{3!(p+1)!}F^{(p+4)}_{ia_0\cdots a_pjk,a}H^{ijk,a}\right)\nonumber
\eeqa
  The  redundant fields $F^{(6)},\cdots, F^{(9)}$ in above action are related to the magnetic dual of the RR field strengths  $F^{(1)},\cdots, F^{(4)}$ as $F^{(10-n)}=*F^{(n)}$ for $n=1,2,3,4$. 
For the self-dual D$_3$-brane in the Einstein frame,  we will show that up to the overall factor of $e^{-\phi}$, the above couplings can be written in $SL(2,R)$ invariant forms. 

The quadratic B-field couplings have been added to  the CS action \reef{CS} by requiring the consistency of this action with the linear T-duality \cite{Becker:2010ij,Garousi:2010rn}. These couplings in the string frame are
\beqa
S_{_{CS}}&\!\!\!\!\!\supset\!\!\!\!\!&\frac{-\pi^2\alpha'^2T_p}{ 2!2!24(p-3)!}\int d^{p+1}x\epsilon^{a_0\cdots a_{p}}{\cal C}^{(p-3)}_{a_4\cdots a_{p-4}}\bigg[\frac{1}{2}H_{a_0a_1a,i}H_{a_2a_3}{}^{a,i}-\frac{1}{2}H_{a_0a_1i,a}H_{a_2a_3}{}^{i,a}
\bigg]+\cdots\labell{Tf41new}
\eeqa
 They have been  confirmed by the S-matrix element of one RR and two NSNS vertex operators \cite{Becker:2010ij,Garousi:2011ut}.  The S-matrix element produces some other couplings   which are refereed by the dots  \cite{Garousi:2011ut}. They would make the couplings to be written in terms of the RR field strength as well.  For the D$_3$-brane case, the S-duality indicates that there should be similar couplings for $CC^{(2)}C^{(2)}$ and $\phi B C^{(2)}$. Including these couplings in the Einstein frame, one can write them in $SL(2,R)$ invariant form up to the overall factor of $e^{-\phi}$. 

The S-duality of the above couplings is  like the S-duality of the $R^4$ corrections to the supergravity action \cite{Green:1981ya,Grisaru:1986dk,Grisaru:1986px,Gross:1986iv}. In the Einstein frame these couplings have an overall factor of $e^{-3\phi/2}$. It has been conjectured in \cite{Green:1997tv} that this factor is in fact the leading order term of the non-holomorphic Eisenstein series $E_s$ with $s=3/2$ at weak coupling. This conjecture has been confirmed with one loop \cite{Green:1997tv} and two loops \cite{D'Hoker:2005jc}. We speculate that the above $O(\alpha'^2)$ corrections to D$_3$-brane  action can be written in $SL(2,Z)$ invariant form by extending the weak coupling factor $e^{-\phi}$ to the regularized non-holomorphic Eisenstein series $E_s$ with $s=1$. This function appears also in the $R^2$ terms of the D$_3$-brane action \cite{Bachas:1999um,Basu:2008gt}.


An outline of the paper is as follows: We begin the section 2 by examining the disk-level S-matrix element of two RR vertex operators from which we find the quadratic RR couplings on the world volume of  D$_p$-branes at order $O(\alpha'^2)$. In section 3, we show that for the self-dual D$_3$-brane case and for the RR two-form,  the couplings are exactly the same as the B-field couplings \reef{LDBI}. We then write the result in $SL(2,Z)$ invariant form using the $SL(2,R)$ matrix $\cM$ which appears  in the Type IIB  supergravity action, and the regularized non-holomorphic Eisenstein series $E_s$ with $s=1$.  In this section we write also the couplings of the  RR four-form and the RR scalar in S-dual form using $E_1(\tau,\bar{\tau})$. In section 4, we write the CS couplings  in $SL(2,Z)$ invariant form. We end this paper by summarizing the new disk level couplings that are predicted by requring the consistency of the DBI action \reef{DBI} and the CS action \reef{CS} with the S-duality.

\section{RR couplings from S-matrix }

 The scattering amplitude of two RR states from D$_p$-brane  is given by \cite{Garousi:1996ad}
 \beqa
 A(\veps_1,p_1;\veps_2,p_2)&=&-\frac{\alpha'^2T_p}{16\times 32}K(1,2)\frac{\Gamma(-\alpha' t/4)\Gamma(\alpha' q^2)}{\Gamma(1-\alpha't/4+\alpha' q^2)}\nonumber\\
 &=&\frac{T_p}{16\times 32}K(1,2)\left(\frac{4}{q^2t}+\z(2)\alpha'^2+O(\alpha'^4)\right)\labell{Amp}
 \eeqa
where $q^2=p_1^ap_1^b\eta_{ab}$ is the momentum flowing along the world-volume of D-brane, and $t=-(p_1+p_2)^2$ is the momentum transfer in the transverse direction. The kinematic factor is
\beqa
K(1,2)&=&\left(2q^2a_1+\frac{t}{2}a_2\right)\labell{kin}
\eeqa
where
\beqa
a_1(n,m,p)&=&-\frac{1}{2}\Tr(P_-\Ga_{1(n)}M_p\ga_{\mu}C^{-1}M_p^T\Ga^T_{2(m)}C\ga^{\mu})
\labell{fintwo}\\
\nonumber\\
a_2(n,m,p)&=&\frac{1}{2}\Tr(P_-\Ga_{1(n)}M_p\ga_{\mu})\Tr(P_-\Ga_{2(m)}M_p\ga^{\mu})
\labell{finthree}\nonumber
\eeqa
 where
\beqa
\Gamma_{\alpha(n)}=\frac{1}{n!}(F_{\alpha})_{\nu_1\cdots\nu_n}\gamma^{\nu_1}\cdots\gamma^{\nu_n}\nonumber\\
M_p=\frac{\pm 1}{(p+1)!}\eps_{a_0\cdots a_p}\gamma^{a_0}\cdots\gamma^{a_p}
\eeqa
where $F_{\alpha}$ for $\alpha=1,2$ is  the linearized RR field strength $n$-form and $\eps$ is the volume $p+1$-form of the $D_p$-brane. In equation \reef{kin}, $P_-=\frac{1}{2}(1-\gamma_{11})$ is the chiral projection operator. The $\gamma_{11}$  gives the magnetic couplings and $1$ gives the electric couplings. The first term in \reef{Amp} produces the massless poles resulting from  the $(\alpha')^0$ order of the DBI and CS couplings on the D-brane, and  the supergravity couplings in the bulk.  The second term in \reef{Amp} should produce $(\alpha')^2$ couplings of two RR fields  on the world volume of D$_p$-brane in which we are interested.

Using various identities, $a_1$ can be simplified for electric components of the RR field strength  to \cite{Garousi:1996ad} 
\beqa
a_1(n,m,p)&=&\frac{8}{n!}\delta_{mn}\bigg[\Tr(D)F_{1(n)}\inn F_{2(n)}-2nD^{\lambda}{}_{\kappa}F_{1\lambda\nu_2\cdots \nu_n} F_2^{\kappa\nu_2\cdots \nu_n}\bigg]
\eeqa
where  the matrix $D^{\mu}_{\nu}$ is diagonal with +1 in the world volume directions and -1 in the transverse directions. The degree of the RR field strength $n$ in $a_1$ is  independent of the dimension of the D$_p$-brane. For the magnetic components, one finds the same result but for different $n$, \ie $n'=10-n$. Using the fact that the redundant field strength $F^{(10-n)}$ for $n=1,2,3,4$ are the magnetic dual of $F^{(n)}$, \ie $F^{(10-n)}=*F^{(n)}$, one finds the following result for the magnetic components of the RR vertex operator:
\beqa
a_1(n',m',p)&=&\frac{8}{n'!}\delta_{m'n'}\bigg[\Tr(D)(*F_{1(n)})\inn (*F_{2(n)})-2n'D^{\lambda}{}_{\kappa}(*F^{(n)})_{1\lambda\nu_2\cdots \nu_{n'}} (*F^{(n)})_2^{\kappa\nu_2\cdots \nu_{n'}}\bigg]\nonumber
\eeqa

The degree of the RR field strength in  $a_2$ depends on the dimension of the D$_p$-brane. To see this consider the factor $\Tr(P_-\Ga_{1(n)}M_p\ga_{\mu})$ in $a_2$. For the  electric components it  is nonzero only for $n=p$ and for $n=p+2$, and for magnetic components it is nonzero for $(10-n)=p$ and for $(10-n)=p+2$. Performing the trace in each case one finds for the electric components
\beqa
\Tr(P_-\Ga_{1(p)}M_p\ga_{\mu})&=&\frac{16}{p!}\delta^{a_0}{}_{\mu}F_1^{a_1\cdots a_p}\eps_{a_0\cdots a_p}\nonumber\\
\Tr(P_-\Ga_{1(p+2)}M_p\ga_{\mu})&=&\frac{16}{(p+1)!}F_1^{a_0\cdots a_p}{}_{\mu}\eps_{a_0\cdots a_p}
\eeqa
We have not pay attention to the signs on the right hand sides because we are interested in $a_2$ which is quadratic multiple of each term. One can easily verify that $a_2$ is nonzero only for the cases 
\beqa
a_2(n=m=p)&=&\frac{8\times 16}{n!}F^{(n)}_{1}\inn V\inn F^{(n)}_{2}\nonumber\\
a_2(n=m=p+2)&=&\frac{8\times 16}{(n-1)!}(F_1^{(n)})^{a_0\cdots a_p}{}_{i}(F_2^{(n)})_{a_0\cdots a_p}{}^{i}\nonumber
\eeqa
where the notation $F^{(n)}_{1}\inn V\inn F^{(n)}_{2}$ means the indices are contracted with the world volume metric $\eta^{ab}$. 

For the magnetic components, one finds
\beqa
a_2(n'=m'=p)&=&\frac{8\times 16}{n'!}(*F_1^{(n)})\inn V\inn (*F_2^{(n)})\nonumber\\
a_2(n'=m'=p+2)&=&\frac{8\times 16}{(n'-1)!}(*F_1^{(n)})^{a_0\cdots a_p}{}_{i}(*F_2^{(n)})_{a_0\cdots a_p}{}^{i}\nonumber
\eeqa
where $n'=10-n$. Again the kinematic factor $a_2$ for the magnetic components is the same as for the electric components but for  $*F$. 

The kinematic factor for the electric RR fields is
\beqa
K(1,2)&=&-\sum_{n}\left(2a_1p_1\inn V\inn p_2+a_2p_1\inn p_2\big[\delta_{n,p}+\delta_{n,p+2}\big]\right)
\eeqa
where the summation is over  $n=1,2,3,4,5$. The  field theory corresponding to this kinematic factor for D$_p$-brane in the string frame is
\beqa
S_{_{DBI}}&\!\!\!\!\supset\!\!\!\!&-\frac{\pi^2\alpha'^2T_p}{48\times 32}\int d^{p+1}x\,e^{\phi}\sqrt{-G}\sum_n\frac{16}{n!}\bigg[(p-4)F^{(n)}{}_{,a}\inn F^{(n)\,,a}-nD^{\mu}{}_{\nu}F^{(n)}{}_{\mu}{}_{,a}\inn F^{(n)\nu}{}^{,a}\nonumber\\
&&+4F^{(n)}{}_{,\mu}\inn V\inn F^{(n)\,,\mu}\delta_{n,p}+4nF^{(n)}{}_{i,\mu}\inn V\inn F^{(n)\,i,\mu}\delta_{n,p+2}\bigg]\labell{FF}
\eeqa
where we have also used the standard convention that the RR fields are rescaled as $C\rightarrow e^{\phi}C$. This is the reason why there is the dilaton factor $e^{\phi}$ in above action  instead of the expected factor of  $e^{-\phi}$ for the disk amplitude. Similar rescaling has been also used in couplings \reef{LCS}. 
The magnetic couplings are the same as above with replacing $F$s with $*F$s. 
The above action gives  all quadratic RR couplings at order $O(\alpha'^2)$ for any D$_p$-brane. However, we are interested in the   world volume couplings   of the self-dual D$_3$-brane.


\section{S-duality of DBI part}

Let us start with the couplings of the two-forms. The couplings of NSNS two-form are given in \reef{LDBI} which have been found in \cite{Garousi:2009dj} by consistency of the curvature couplings with T-duality. Compatibility of these couplings with S-duality requires similar couplings for the RR two-form. Writing the spacetime indices in \reef{FF} in terms of the world volume and the transverse indices, one finds that for D$_3$-brane and for RR two-form the above action simplifies to 
\beqa
S_{_{DBI}}&\!\!\!\!\supset\!\!\!\!&-\frac{\pi^2\alpha'^2T_{3}}{48}\int d^{4}x\,e^{\phi}\sqrt{-G}\bigg[\frac{1}{6}F_{ijk,a}F^{ijk,a}+\frac{1}{3}F_{abc,i}F^{abc,i}-\frac{1}{2}F_{bci,a}F^{bci,a}
\bigg]\labell{FF1}
\eeqa
These couplings are similar to the B-field couplings in \reef{LDBI}. 

To study the transformation of the above couplings under S-duality, one should first rescale the metric from string frame to the Einstein frame $G_{\mu\nu}=e^{\phi/2}g_{\mu\nu}$. The B-field couplings \reef{LDBI} are multiplied by  $e^{-2\phi}$ and  the dilaton drops out of the above RR couplings. The D$_3$-brane and the Einstein frame metric are invariant under $SL(2,Z)$, and the B-field and the RR two-form transform as doublet, \ie
\beqa
{\cal B}^{(2)}=\pmatrix{B^{(2)} \cr 
C^{(2)}}
\eeqa
Under a $SL(2,Z)$ transformation by
\beqa
\Lambda=\pmatrix{d&c \cr 
b&a}
\eeqa
the ${\cal B}$-field  transforms linearly by the rule 
\beqa
 {\cal B}\rightarrow \Lambda {\cal B}
 \eeqa
The axion and the dilaton combine  into a complex scalar field $\tau=C+ie^{-\phi}$. This field  transforms as
\beqa
\tau\rightarrow \frac{a\tau+b}{c\tau +d} 
\eeqa
Now consider  the matrix ${\cal M}$ 
 \beqa
 {\cal M}=e^{\phi}\pmatrix{|\tau|^2&-C \cr 
-C&1}\labell{M}
\eeqa
which transforms under the $SL(2,Z)$ as
\beqa
{\cal M}\rightarrow (\Lambda ^{-1})^T{\cal M}\Lambda ^{-1}
\eeqa
This matrix appears in  the $SL(2,Z)$ form of the type IIB supergravity. 
Using this matrix, one can rewrite the couplings \reef{LDBI} and \reef{FF1}  in the Einstein frame as:
\beqa
S_{_{DBI}}&\!\!\!\!\!\!\supset\!\!\!\!\!\!&-\frac{\pi^2\alpha'^2T_{3}}{48}\int d^{4}x\,e^{-\phi}\sqrt{-g}\bigg[\frac{1}{6}{\cal F}^T_{ijk,a}\cM\cF^{ijk,a}+\frac{1}{3}\cF^T_{abc,i}\cM\cF^{abc,i}-\frac{1}{2}\cF^T_{bci,a}\cM\cF^{bci,a}
\bigg]\labell{FF2}
\eeqa
where $\cF=d\cB$. Apart from the overall dilaton factor $e^{-\phi}$, it is invariant under the $SL(2,Z)$ transformation. 


The above situation  is like the $R^4$ corrections to the supergravity action which apart from the factor $\z(3)e^{-3\phi/2}$, the couplings are invariant under the $SL(2,Z)$ transformation. In that case,  the tree level,   one loop  and one-instanton results are the three leading order terms of the non-holomorphic Eisenstein series $E_s$ with $s=3/2$. So it has been conjectured in \cite{Green:1997tv} that the $SL(2,Z)$ invariant coupling is $E_{3/2}R^4$. In particular, this conjecture indicates that there is no perturbative corrections to $R^4$ other than one loop. It has been shown in \cite{D'Hoker:2005jc} that there is no two loop correction to this action.

For general $s$, this series is defined by 
\beqa
2\z(2s)E_s(\tau,\bar{\tau})&=&\sum_{(m,n)\neq(0,0)}\frac{\tau_2^s}{|m+n\tau|^{2s}}\labell{series}
\eeqa
where $\tau=\tau_1+i\tau_2$. It is invariant under the $SL(2,Z)$ transformation. This function satisfies the following differential equation:
\beqa
4\tau_2^2\prt_{\tau}\prt_{\bar{\tau}}E_s&=&s(s-1)E_s
\eeqa
This equation has two solutions $\tau_2^s$ and $\tau_2^{1-s}$ corresponding to two particular orders of perturbation theory, and infinite number of non-perturbative solutions. For $s=1$, however, the series \reef{series} diverges logarithmically. The regularized function has  the following expansion \cite{Bachas:1999um,Basu:2008gt}:
\beqa
2\z(2)E_1(\tau,\bar{\tau})&=&\z(2)\tau_2-\frac{\pi}{2}\ln(\tau_2)+\pi\sqrt{\tau_2}\sum_{m\neq 0,n\neq 0}\left|\frac{m}{n}\right|^{1/2}K_{1/2}(2\pi|mn|\tau_2)e^{2\pi imn\tau_1}\labell{regE}
\eeqa
where the first term  corresponds to $n=0$ in the series \reef{series}. This term is exactly the dilaton factor in \reef{FF2}. This may indicate that the factor $e^{-\phi}$ in \reef{FF2} should be replaced by the $SL(,Z)$ invariant function $E_1$.
 
Another  evidence for this replacement is the following. The B-field couplings \reef{LDBI} are related to the gravity couplings \reef{DBI} by T-duality \cite{Garousi:2009dj}. On the other hand the $SL(2,Z)$ invariant form of the gravity couplings has this nonperturbative factor \cite{Bachas:1999um,Basu:2008gt}. In fact, the regularized  function \reef{regE} is proportional to $\log(\tau_2|\eta(\tau)|^4)$  \cite{Bachas:1999um} where $\eta(\tau)$ is the Dedekind $\eta$-function. The $\log\tau_2$ piece is nonanalytic and comes from the annulus   \cite{Basu:2008gt} and the remaining part appears in the Wilsonian effective action found in \cite{Bachas:1999um}. Hence, one expects the $SL(2,Z)$ invariant form of the  $\cB$-field couplings \reef{FF2}  to be
\beqa
S_{_{DBI}}&\!\!\!\!\supset\!\!\!\!&-\frac{\pi^2\alpha'^2T_{3}}{24}\int d^{4}x\,E_1(\tau,\bar{\tau})\sqrt{-g}\bigg[\frac{1}{6}{\cal F}^T_{ijk,a}\cM\cF^{ijk,a}\nonumber\\
&&\qquad\qquad\qquad+\frac{1}{3}\cF^T_{abc,i}\cM\cF^{abc,i}-\frac{1}{2}\cF^T_{bci,a}\cM\cF^{bci,a}
\bigg]\labell{FF3}
\eeqa
The second term in the expansion of $E_1$ is the annulus contribution to the 1PI effective action.  All other terms are D-instanton contributions.

The next case that we consider is the  couplings of two $F^{(5)}$ on the world volume of D$_3$-brane. The RR potential $C^{(4)}$ is invariant under the $SL(2,Z)$ transformation. One can easily confirm that the couplings \reef{FF} for $C^{(4)}$ in the Einstein frame  have the dilaton factor $e^{-\phi}$. Replacing this factor with the $SL(2,Z)$ invariant function $E_1(\tau,\bar{\tau})$, one finds the following S-dual couplings:
\beqa
S_{_{DBI}}&\!\!\!\!\supset\!\!\!\!&-\frac{\pi^2\alpha'^2T_3}{48\times 5!}\int d^{4}x\,E_1(\tau,\bar{\tau})\sqrt{-g}\bigg[-F^{(5)}{}_{,a}\inn F^{(5)\,,a}-5D^{\mu}{}_{\nu}F^{(5)}{}_{\mu}{}_{,a}\inn F^{(5)\nu}{}^{,a}\nonumber\\
&&+20F^{(5)}{}_{i,\mu}\inn V\inn F^{(5)\,i,\mu}\bigg]\labell{FF5}
\eeqa
where we have also included the  $*F^{(5)}$ terms. Similar couplings as those in the  first line above,  without the factor $E_1$,  can be written for the $F^{(5)}$ couplings on the world volume of  D$_7$-brane. Since it is S-duality invariant, the $F^{(5)}$ couplings in \reef{FF}  are the couplings for all $(p,q)$ 7-branes.

The disk level S-matrix element of two NSNS vertex operators shows that there is no couplings for one dilaton and one graviton at order $O(\alpha'^2)$ for D$_3$-brane \cite{Garousi:2009dj}. The kinematic factor  for D$_3$-brane is
\beqa
K(\phi, \veps)&=&\Tr(\veps)\left(\frac{4q^4}{\sqrt{8}}\right)\labell{phih}
\eeqa
where $\veps$ is the polarization of the symmetric tensor. Using the traceless of the graviton polarization, one finds the above kinematic factor is zero for graviton\footnote{This term which is zero for one graviton and one dilaton, did not considered in \cite{Garousi:2009dj}. This term has non-zero contribution for the kinematic factor of two diatons.}. This is consistent with S-duality because RR couplings \reef{LCS}  have no  coupling between one graviton and one axion. It is also consistent with the couplings \reef{DBI2}. Using various on-shell relations, one can show that in the Einstein frame these couplings become
\beqa
S_{_{DBI}}&\!\!\!\!\supset\!\!\!\!&\frac{\pi^2\alpha'^2T_{p}}{48}\int d^{p+1}x\,e^{-\phi}\sqrt{-g}\bigg[R_{abcd}R^{abcd}-2\hR_{ab}\hR^{ab}
-R_{abij}R^{abij}+2\hR_{ij}\hR^{ij}\nonumber\\
&&-(p-3)\big[\hR_{ab}\phi^{,ab}-\hR_{ij}\phi^{,ij}\big]-\frac{(p-3)^2}{8}\big[\phi_{,ab}\phi^{,ab}-\phi_{,ij}\phi^{,ij}]+\phi_{,ab}\phi^{,ab}\bigg]\labell{DBI3}
\eeqa
which gives zero coupling for the linear dilaton for the case $p=3$. Replacing the polarization tensor $\veps$ in \reef{phih} with the dilaton polarization, one finds the last term in the above equation. 

 Under the S-duality both the dilaton and the RR scalar appears in the complex field $\tau$, so one  expects that there should be similar coupling  for the RR scalar at order $O(\alpha'^2)$ for D$_3$-brane.  The couplings \reef{FF} for the RR scalar for the D$_3$-brane in the Einstein frame become
\beqa
S_{_{DBI}}&\!\!\!\!\supset\!\!\!\!&\frac{\pi^2\alpha'^2T_3}{48}\int d^{4}x\,e^{\phi}\sqrt{-g}C_{,ab}C^{,ab}\labell{FF6}
\eeqa
To study the S-duality of this coupling we add the dilaton coupling in \reef{DBI3} to the above equation, \ie
\beqa
S_{_{DBI}}&\!\!\!\!\supset\!\!\!\!&\frac{\pi^2\alpha'^2T_3}{48}\int d^{4}x\,e^{-\phi}\sqrt{-g}\bigg[\phi_{,ab}\phi^{,ab}+e^{2\phi}C_{,ab}C^{,ab}\bigg]\labell{FF611}
\eeqa
The terms in the bracket can be extended to the $SL(2,Z)$ invariant form  by  including the disk-level nonlinear terms, and the dilaton factor can be extended to the $SL(2,Z)$ invariant form  by including the annulus and the D-instanton effects.
The S-dual extension of the coupling \reef{FF611} is 
\beqa
S_{_{DBI}}&\!\!\!\!\supset\!\!\!\!&-\frac{\pi^2\alpha'^2T_3}{96}\int d^{4}x\,E_1(\tau,\bar{\tau})\sqrt{-g}\Tr(\cM_{,ab}(\cM^{-1})^{,ab})\labell{FF61}
\eeqa
where the matrix $\cM$ is the one appears in \reef{M}

\section{S-duality of CS part}

The linear NSNS couplings in the CS action are given in \reef{LCS}. To study the S-duality of these couplings for the self-dual D$_3$-brane, we begin  by witting the couplings in the first line in the Einstein frame, \ie
\beqa
S_{_{CS}}&\!\!\!\supset\!\!\!&-\frac{\pi^2\alpha'^2T_3}{24\times 2!2!}\int d^{4}x\,\eps^{a_0\cdots a_3}e^{-\phi}\bigg[{ F}_{ia_2a_3,a}H_{a_0a_1}{}^{a,i}-{ F}_{aa_2 a_3,i}H_{a_0a_1}{}^{i,a}\bigg]\labell{LCS1}
\eeqa
To write it in a $SL(2,Z)$ invariant form, we first introduce the $SL(2,Z)$ matrix
\beqa
\cN=\pmatrix{0&1 \cr 
-1&0}
\eeqa
This matrix has the property 
\beqa
\cN=(\Lambda ^{-1})^T{\cal N}\Lambda ^{-1}
\eeqa
 Using this matrix, one can rewrite the terms in the parenthesis in the S-dual form $\cF^T\cN\cF$. Then using the same idea as in previous section, one can extend it to the following $SL(2,Z)$ invariant form:
\beqa
S_{_{CS}}&\!\!\!\supset\!\!\!&-\frac{\pi^2\alpha'^2T_3}{12\times 2!2!}\int d^{4}x\,\eps^{a_0\cdots a_3}E_1(\tau,\bar{\tau}){ \cF^T}_{a_0 a_1a,i}\cN \cF_{a_2a_3}{}^{i,a}\labell{LCS2}
\eeqa
The second term in the expansion of $E_1(\tau,\bar{\tau})$ can be calculated with the annulus level S-matrix element of one RR and one NSNS vertex operators.

The  RR couplings in the second line of \reef{LCS} for the D$_3$-brane  in the Einstein frame are proportional to the dilaton factor $e^{-\phi}$. In particular, the term $ \hat{\cR}^{ij}-\phi^{,ij}$ becomes proportional to $\hat{\cR}^{ij}+\eta^{ij}\phi_{,a}{}^a/4$ in the Einstein frame. The last term gives $F^{(5)}_{a_0\cdots a_3i}{}^{,i}$ which is zero on-shell, \ie $F^{(5)}_{a_0\cdots a_3i}{}^{,i}=-F^{(5)}_{a_0\cdots a_3a}{}^{,a}$ where two world volume indices in $F^{(5)}_{a_0\cdots a_3a}$ is identical, hence, it is zero. Using the fact that the Einstein frame metric and $C^{(4)}$ are invariant under the S-duality, and replacing  $e^{-\phi}$ with $E_1(\tau,\bar{\tau})$, one finds the following S-dual couplings:
\beqa
S_{_{CS}}&\!\!\!\supset\!\!\!&-\frac{\pi^2\alpha'^2T_3}{6}\int d^{4}x\,\eps^{a_0\cdots a_3}E_1(\tau,\bar{\tau})\bigg[\frac{1}{2!3!}F^{(5)}_{ia_1\cdots a_3j,a}\cR^a{}_{a_0}{}^{ij}-\frac{1}{4!}F^{(5)}_{a_0\cdots a_3j,i}\hat{\cR}^{ij}\bigg]\nonumber
\eeqa
Note that the above couplings are not invariant under  the standard form of T-duality transformations. This is because the standard T-duality  rules are in string frame whereas the above S-dual couplings are in  the Einstein frame. To study the T-duality of the above couplings one should first write the T-duality transformations in the Einstein frame and then applied to the above action. 

The coupling in the third line of \reef{LCS} for D$_3$-brane is
\beqa
S_{_{CS}}&\!\!\!\supset\!\!\!&\frac{\pi^2\alpha'^2T_3}{24}\int d^{4}x\,\eps^{a_0\cdots a_3}\left(\frac{1}{3!4!}F^{(7)}_{ia_0\cdots a_3jk,a}H^{ijk,a}\right)\labell{FH}
\eeqa
Using the fact that the redundant RR field strength $F^{(7)}$ is the magnetic dual of $F^{(3)}$, \ie $F^{(7)}=*F^{(3)}$, and the relation
\beqa
(*F^{(n)})_{\mu_{n+1}\cdots \mu_{10}}=\frac{1}{n!}\sqrt{-G}\eps_{\mu_1\cdots \mu_{10}}(F^{(n)})^{\mu_1\cdots \mu_n}
\eeqa
one finds that in the Einstein frame again the coupling \reef{FH} has the dilaton factor $e^{-\phi}$. Moreover, the coupling is antisymmetric under changing the RR two form with the B-field. So the coupling \reef{FH} can be written in the Einstein frame as
\beqa
S_{_{CS}}&\!\!\!\supset\!\!\!&\frac{\pi^2\alpha'^2T_3}{24\times 2}\int d^{4}x\,\eps^{a_0\cdots a_3}e^{-\phi}\left(\frac{1}{3!4!}(*F^{(3)})_{ia_0\cdots a_3jk,a}H^{ijk,a}-\frac{1}{3!4!}(*H)_{ia_0\cdots a_3jk,a}(F^{(3)})^{ijk,a}\right)\nonumber
\eeqa
Defining the magnetic dual of ${\cal F}^{(3)}=d\cB^{(2)}$ as
\beqa
{\cal F}^{(7)}=*{\cal F}^{(3)}=\pmatrix{*H^{(3)} \cr 
*F^{(3)}}\, ,
\eeqa
 one finds the following S-dual coupling:
\beqa
S_{_{CS}}&\!\!\!\supset\!\!\!&-\frac{\pi^2\alpha'^2T_3}{24\times 3!4!}\int d^{4}x\,\eps^{a_0\cdots a_3}E_1(\tau,\bar{\tau})({\cal F}^{(7)})^T_{a_0\cdots a_3ijk,a}\cN {\cal F}^{ijk,a}\labell{FH2}
\eeqa
Similar coupling as above,  without the factor $E_1$,  can be written for the couplings in the first line of \reef{LCS} for D$_7$-brane. Therefore,  the couplings in the first line of \reef{LCS} are the couplings for all $(p,q)$ 7-branes.

Finally we consider the Chern-Simons couplings.  Compatibility of the CS action \reef{CS} with T-duality requires the couplings \reef{Tf41new}. We now try to write these B-field couplings in S-dual invariant form. The couplings  \reef{Tf41new} for the self-dual D$_3$-brane   in the Einstein frame are
\beqa
S_{_{CS}}&\!\!\!\supset\!\!\!&-\frac{\pi^2\alpha'^2T_3}{ 2!2!48}\int d^{4}x\epsilon^{a_0\cdots a_{3}} e^{-\phi}\delta C\bigg[H_{a_0a_1a,i}H_{a_2a_3}{}^{a,i}-H_{a_0a_1i,a}H_{a_2a_3}{}^{i,a}
\bigg]+\cdots\labell{HH}
\eeqa
The above couplings have been also confirmed with scattering calculation \cite{Becker:2010ij,Garousi:2011ut}. In above equation we have replaced the RR scalar by $\delta C$ to specify that it is quantum fluctuation. In fact we expect from the couplings \reef{FF3} that the above action to be zero if the RR scalar is a constant background. The compatibility of these couplings with the S-duality indicates that  each $\delta C HH$ coupling to be extended to the following $SL(2,R)$ invariant form:
\beqa
\delta C H_{a_0a_1a,i}H_{a_2a_3}{}^{a,i}\rightarrow \cF^T_{a_0a_1a,i}\delta \cM \cN^{-1}\cM \cF_{a_2a_3}{}^{a,i}\labell{CH1}
\eeqa
The overall   dilaton factor should also be extended to the Eisenstein series $E_1$ by including the one-loop couplings and the D-instanton effects.

For D$_7$-brane, the couplings in the Einstein frame are 
\beqa
S_{_{CS}}&\!\!\!\supset\!\!\!&-\frac{\pi^2\alpha'^2T_7}{  2!2!4!48}\int d^{8}x\epsilon^{a_0\cdots a_{7}} e^{-\phi} C^{(4)}_{a_4\cdots a_7}\bigg[H_{a_0a_1a,i}H_{a_2a_3}{}^{a,i}-H_{a_0a_1i,a}H_{a_2a_3}{}^{i,a}
\bigg]+\cdots\labell{HH11}
\eeqa
The RR four-form is invariant under the S-duality. The couplings $e^{-\phi}HH$ may be extended to the following $SL(2,R)$ invariant form:
\beqa
e^{-\phi} H_{a_0a_1a,i}H_{a_2a_3}{}^{a,i}\rightarrow \cF^T_{a_0a_1a,i} \cM \cF_{a_2a_3}{}^{a,i}\labell{CH2}
\eeqa
Then the resulting $SL(2,R)$ invariant action describes the appropriate couplings for all $(p,q)$ 7-branes.
This ends our illustration of the consistency of the D-brane action at order $O(\alpha'^2)$ with S-duality. 

The consistency of the DBI action \reef{DBI} with the S-duality predicts the following couplings in the Einstein frame at the disk level:
\beqa
S_{_{DBI}}&\!\!\!\supset\!\!\!&-\frac{\pi^2\alpha'^2T_{3}}{48}\int d^{4}x\sqrt{-g}C^2\bigg[\frac{1}{6} H_{ijk,a}H^{ijk,a}+\frac{1}{3}H_{abc,i}H^{abc,i}-\frac{1}{2}H_{bci,a}H^{bci,a}
\bigg]\\
S_{_{DBI}}&\!\!\!\!\!\!\supset\!\!\!\!\!\!&\frac{\pi^2\alpha'^2T_{3}}{24}\int d^{4}x\sqrt{-g}C\bigg[\frac{1}{6} H_{ijk,a}F^{ijk,a}+\frac{1}{3}H_{abc,i}F^{abc,i}-\frac{1}{2}H_{bci,a}F^{bci,a}\nonumber
\bigg]
\eeqa
 and well as the new couplings in \reef{FF61}, \reef{CH1} and \reef{CH2} that result from writing  the $SL(2,R)$ invariants in terms of their components.
Moreover, since the modular function $E_1(\tau,\bar{\tau})$ has annulus contribution, the S-duality also predicts the annulus level couplings for  the S-duality invariant  couplings that we have found, \eg \reef{FF3}. It would be interesting to confirm these couplings by direct calculations.

{\bf Acknowledgments}:  I would like to thank Michael B. Green for useful  discussions. This work is supported by Ferdowsi University of Mashhad under grant p/17212(1389/12/24). 
\bibliographystyle{/Users/Nick/utphys} 
\bibliographystyle{utphys} \bibliography{hyperrefs-final}


\providecommand{\href}[2]{#2}\begingroup\raggedright

\endgroup

\end{document}